\documentstyle[11pt,psfig,aaspp4]{article} 
\def\etal{{\it et al. }}
\def\kms{km~s$^{-1}$~}
\def\vpec{$V_{pec}$~}
\def\hnot{$H_\circ$~}
\def\kmsm{km s$^{-1}$ Mpc$^{-1}$~}

\def\W50{W$_{50}$~}
\font\eightrm=cmr8
\font\eightbf=cmbx8
\font\eightit=cmti8
\font\eightsl=cmsl8
\font\eightmus=cmmi8
\def\smalltype{\let\rm=\eightrm \let\bf=\eightbf
  \let\it=\eightit \let\sl=\eightsl \let\mus=\eightmus
  \baselineskip=9.5pt minus .75pt
  \rm}
\def\medtype{\let\rm=\tenrm \let\bf=\tenbf
  \let\it=\tenit \let\sl=\tensl \let\mus=\tenmus
  \baselineskip=12.5pt minus .75pt
  \rm}

\journalid{337}{Someday 1999}
\articleid{11}{14}

\begin{document}

\noindent{\smalltype {To appear in
{\it The Extragalactic Distance Scale}, M. Livio, M. Donahue and N. Panagia, eds., Cambridge}}

\title{The I--Band Tully--Fisher Relation and the Hubble Constant}

\author {Riccardo Giovanelli}
\affil{Department of Astronomy
and National Astronomy and Ionosphere Center\altaffilmark{1},
Cornell University, Ithaca, NY 14953}

\altaffiltext{1}{The National Astronomy and Ionosphere Center is
operated by Cornell University under a cooperative agreement with the
National Science Foundation.}

\hsize 6.5 truein
\begin{abstract}

The application of the I band Tully--Fisher relation towards
determining the Hubble constant is reviewed, with particular
attention to the impact of scatter and bias corrections in
the relation. A template relation is derived from galaxies in
24 clusters. A subset of 14 clusters with $cz \sim 4000$ to 9000 \kms 
is used as an inertial frame to define the velocity zero point of
the relation. Twelve galaxies with Cepheid distances are used
to establish the absolute magnitude scale of the Tully--Fisher
relation, and thus to determine a value of \hnot $= 70\pm5$ \kmsm.
Estimates of the peculiar velocities of the Virgo and Fornax
clusters are also given. Assuming that the distance to Fornax
is 18.2 Mpc (N1365), \hnot $=76\pm8$ \kmsm. Assuming that Virgo
lies at 17.4 Mpc (M100, N4496, N4639), \hnot $=67\pm8$ \kmsm.

\end{abstract}

\section{Introduction}

The luminosity--linewidth relation has become the most widely used 
technique for the determination of redshift--independent galaxy distances. 
While the relationship between kinematic and photometric parameters 
of spirals was investigated early on by Roberts (1969, 1975), 
and Ernst Oepik applied it to derive a distance of 450 kpc 
to M31 in 1922 --- a much closer value to the true one than those around 
200--250 kpc commonly adopted until the 1950's ---, it is the merit of R.B. 
Tully and J.R. Fisher (1977) that of forcefully proposing the technique and 
underscoring its great potential for extragalactic astronomy and cosmology. 

The Tully--Fisher (TF) relation was first obtained by correlating blue 
luminosity and the velocity width of the 21cm line, as observed with 
telescopes that yield the integrated spectrum of the galaxy, rather than
a map of the velocity field. Since the technique requires that target 
galaxies be fairly inclined in order to minimize the amplitude of the
corrections needed to recover the disk's rotational speed, it was realized 
early that uncertain corrections for internal extinction of the stellar 
flux would impose a serious limitation on the method. Aaronson, Huchra and 
Mould (1980) adopted a photometric datum shifted to the H band, 
thus drastically reducing the amplitude of, and the
error on internal extinction corrections. In the late 1980's, the advent of 
CCD devices stimulated the adoption of I and R band as the bandpasses of 
choice for TF work. Sky background at those wavelengths is relatively 
low (as compared to H and K bands), detectors have high efficiency and 
large fields and data acquisition is relatively fast even with small 
aperture telescopes. The population contributing most of the light at I 
band is comprised of solar--like stars, several Gyr old. Thus disks are 
well outlined but of smoother appearance than seen in the bluer parts of
the visible spectrum, and their apparent inclinations to the line of sight 
can be reliably determined. Moreover, processes operating in clusters that 
may alter the star formation rate in galaxies will have a retarded effect 
on the red and infrared light of disks; thus, smaller --- if any --- 
systematic differences are expected between the TF relation of cluster 
and that of field galaxies, if I or R band photometry is used (Pierce and 
Tully 1992). HI single--dish linewidths sample effectively the outer
regions of disks and yield values close to twice the asymptotic value of 
the rotational speed. They are also impervious to uncertainties on major 
axis position angle. However, not all sky is reachable by 
large aperture radio telescopes, and velocity widths 
have also been extensively derived from single--slit spectra, principally 
targeted at nebular H$_\alpha$, N and S lines in the red part of the 
spectrum, and to a lesser extent from  H$_\alpha$ Fabry--Perot 
imaging spectroscopy. HI synthesis imaging has so far had a small impact 
on this field of work, although the advent of broad--band spectrographs 
in arrays may make cluster TF work an advantageous proposition. 

Much observational and theoretical work has been carried out on the TF 
relation, and a fair review of even the main individual works is impossible 
within the space constraints of this presentation. Recent CCD TF work
includes the surveys of Pierce and Tully (1988, 1992), Han and Mould (1992),
Schommer \etal (1993), Mathewson \etal (1992), Courteau (1992), Willick 
(1990), Bernstein \etal (1994), Bureau \etal (1996) and Giovanelli \etal 
(1996, hereafter G96). By mid--1996, the TF distances of a few thousand 
galaxies had been estimated, principally with the purpose of determining 
the characteristics of the peculiar velocity field within $cz \sim 10,000$ 
\kms. 

The investigation of the peculiar velocity 
($V_{pec}$) field using the TF technique does {\it not} require an 
absolute distance calibration of the luminosity--width relation. The
observed radial velocity of a galaxy at distance $d$ is
$$cz = H_\circ d + [{\bf V}_{pec}({\bf d}) -  {\bf V}_{pec}(0)]\cdot 
({\bf d}/d) \eqno (1)$$
where {\bf V}$_{pec}$ is the peculiar velocity vector and {\bf V}$_{pec}(0)$
its value at the observer's location. If the CMB dipole (Kogut \etal 1993)
is interpreted
as a Doppler shift, then in the CMB reference frame {\bf V}$_{pec}(0) = 0$.
TF yields the distance $H_\circ d$ in \kms. Thus, a template with accurate 
slope and zero--point is required, such 
that any galaxy falling on the template can be considered at rest with 
respect to the comoving reference frame. A {\it velocity calibration} 
of the TF zero--point is necessary, which is introduced by assuming that 
the TF relation of one cluster, or the composite of a set of clusters 
define \vpec = 0. Section 4 deals with the process of establishing such 
calibration. Once a reliable TF template is in hand, its luminosity scale 
can be absolutely calibrated by inspecting the 
location in the TF plane of galaxies with known distances, as obtained 
via primary indicators. This is equivalent to estimating \hnot.

In this presentation, the physical basis of the TF relation will be 
reviewed in section 2, in terms of dynamical and scaling arguments. 
In section 3, the important issue of the scatter about the mean TF 
relation will be discussed: on it hinge not only the limits 
of applicability of the technique, but also the impact of bias. Differences
in estimate of the latter are largely responsible for wide discrepancies 
in inferences of both the value of the Hubble constant and the amplitude 
of peculiar velocities, while the characteristics of scatter affect the
issue of whether inverse formulations of the TF relation are bias--free 
tools. In section 4, the derivation of a template TF relation, from a 
sample of clusters spread between $cz\sim 1000$ and 10,000 \kms, will 
be presented, with emphasis on the treatment of the correction for bias. 
In sections 5--7, the TF template 
relation (a) will be calibrated using a set of nearby calibrators with 
available Cepheid distances and (b) will be used to estimate the Virgo
and Fornax clusters' peculiar velocities. These determinations
will be used to infer values of H$_\circ$. Throughout this paper,
the parametrization H$_\circ = 100 h$ \kmsm will be used.

\section {The Physical Basis of the TF Relation}

There is no detailed physical understanding of the TF relation, the
interpretation of which relies on simple scaling relations
and dynamical arguments. 
In the scenario most recently discussed by Eisenstein and Loeb (1996),
consider a galaxy of mass $M$ collapsing from a spherical cloud at epoch 
$t_{coll}$; the turnaround radius $R_{ta}$ at that epoch is 
$$R_{ta} \propto M^{1/3} t_{coll}^{2/3} \eqno (2)$$
according to the spherical collapse model of Gunn and Gott (1972) for
$\Lambda = 0$ cosmologies. The total energy of the object after
virialization is 
$$E\propto M\sigma^2 \propto G M^2/R_{ta} \eqno (3)$$
where $\sigma$ is its velocity dispersion.
Assuming that a galaxy results from collapse at a single epoch and not 
from a succession of mergers, the combination of (2) and (3) yields 
$$\sigma \propto (M/t_{coll})^{1/3}, \eqno (4)$$
so if all galaxies collapse in single events at the same epoch,
and mass--to--light ratios don't vary significantly
$L \propto \sigma^3$.
Variations in the formation history of galaxy systems should be expected
to introduce substantial scatter in the relation, as will be discussed 
in section 3.

Invoking an alternative set of scaling relations,
consider a pure exponential disk of central
disk surface brightness $I(0)$ and scale length $r_d$; its total
luminosity is 
$$L_d \propto r_d^2 I(0) \eqno (5)$$ 
On the other hand, the mass internal to radius $R$ is $M(R) \propto R V^2$,
and if the rotation curve flattens in the outer regions of the
disk as is usually the case for spiral galaxies, the total mass is
$$M_{tot} \propto r_d V_{max}^2 \eqno (6)$$
Combining eqns. (5) and (6), we can write
$$L_d \propto (M_{tot}/L_d)^{-2} V_{max}^4/I(0) \eqno (7)$$
If a dark matter halo is present so that the disk mass is 
$M_{d} = \Gamma M_{tot}$,
$$L_d \propto (M_d/L_d)^{-2} \Gamma^2 V_{max}^4/I(0) \eqno (8)$$
When a number of ``standard'' assumptions are made, i.e. that $\Gamma \sim $ const,
$M_d/L_d \sim$ const and $I(0) \sim$ const (Freeman's law, 1970),
$L_d \propto V_{max}^4$, which resembles the TF relation. In practice,
none of the assumptions of constancy for $M_d/L_d$, $\Gamma$ and $I(0)$
apply; all those parameters exhibit mild dependencies on $V_{max}$ 
(or $L_d$), reducing the exponent to values $n<4$, in a measure that
depends on the adopted photometric band. 

Empirical calibrations of the TF relation yield power law
behavior as 
$$L_d \propto V_{max}^n, \eqno(9)$$ 
with values of $n\sim 3$ in the I band.
Some workers (e.g. Aaronson et al. 1986; Willick 1990;
Pierce and Tully 1996) have found significant departures from a single 
power law behavior, and quadratic or bilinear TF fits have been adopted.
It has been argued that inappropriate extinction corrections (Giovanelli 
et al. 1995) or samples including a mixture of morphological types 
(G96) may result in TF departures from linearity. 
There is however no {\it a priori} reason to expect that the TF 
relation be strictly linear.

\section {The Scatter About the TF Relation}

In order to correctly infer the predictive power of the TF method
and to adequately estimate the amplitude of its biases,
a fair understanding of the nature of the associated scatter and 
its sources is necessary.

The scatter in the TF relation arises from several sources: errors in 
the {\it measurement} of TF parameters and uncertainties associated with the 
{\it corrections} applied to them combine with {\it variance in the
galactic properties} produced by different formation and evolutive
histories, characteristic of each object. The latter is often referred 
to as the {\it intrinsic} contribution to scatter, which may appear
in the form of velocity field distortions, deviations from disk 
planarity, other gravitational and disk asymmetries, etc. 
Several misconceptions regarding the nature of the TF scatter often
appear in the literature: that it is well represented by a single number; 
that the measurement and correction errors fully account for the observed 
dispersion; that only errors on the velocity widths are important.
G96 have made a detailed appraisal of the sources of I band TF scatter. 
Their principal results, illustrated in Figure 1, are:
\begin{figure} 
\centerline{\psfig{figure=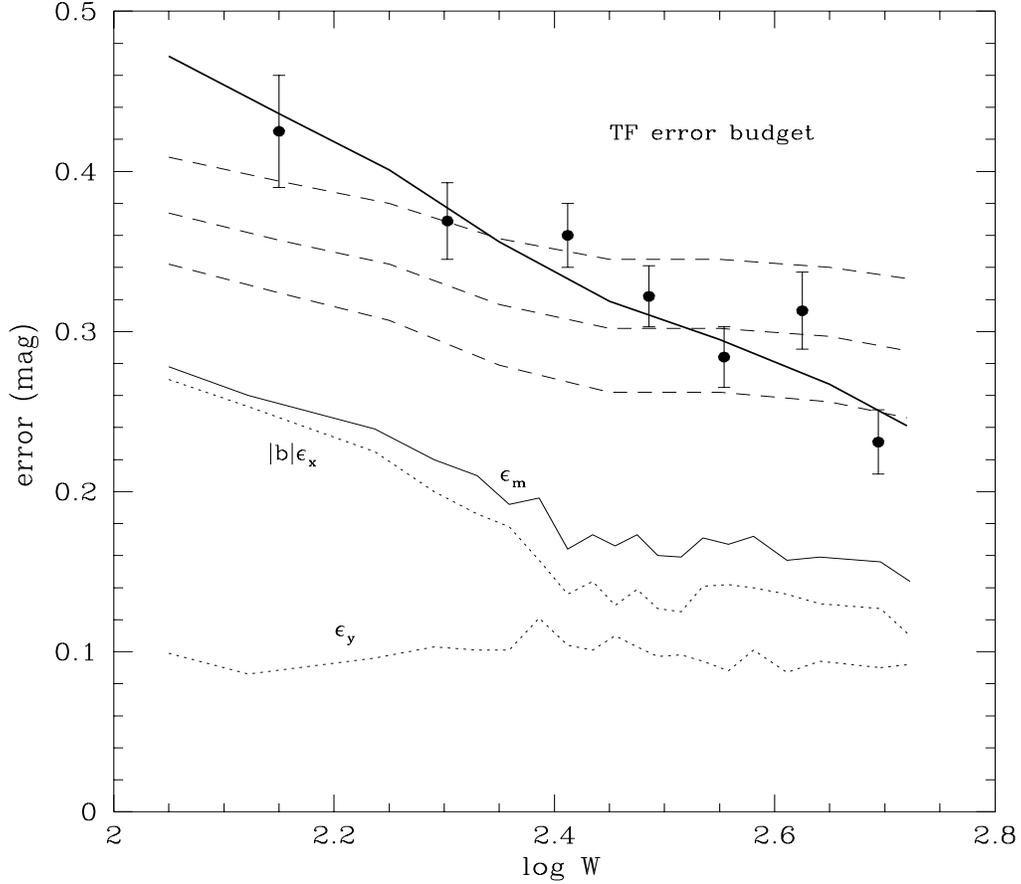,height=5.0in,width=5.5in}}
\caption {Error budget in the TF relation. Mean errors in the 
measurements and processing of $\log W$ and magnitudes are plotted
as dotted lines and respectivelly labelled $\epsilon_x$ and $\epsilon_y$.
The total measurement plus processing mean error is labelled $\epsilon_m$.
The three dashed lines correspond to the addition in quadrature of the
smoothed curve given by $\epsilon_m$ and, respectively bottom to top,
an instrinsic scatter contribution of 0.20, 0.25 and 0.30 mag. The
observed mean r.m.s. scatter is given by the filled symbols and the
solid line is a model of the total scatter as given by G96.
}
\end{figure}

\noindent (i) {\it The total TF scatter cannot be represented by a
single value}; rather, {\it it varies monotonically with the velocity
width (luminosity) of the galaxy}, as shown by the solid symbols and
the heavy line in Figure 1. Over the range of velocities generally
used for TF studies, the r.m.s. scatter in magnitudes varies by a 
factor of $\sim 2$, between 0.5 and 0.25 mag (which translate in
distance errors of respectively 25\% and 12\%). The use of low width
objects is not particularly advantageous (see Hoffman \& Salpeter 1996). 
Rather, it is the high width galaxies that generally yield the highest 
accuracy.

\noindent (ii) {\it Measurement and processing errors are important
contributors to, but they cannot fully account for the amplitude of 
the total error}. Average errors on the magnitude and on the widths 
are shown in Figure 1 as dotted lines (that on the width is multiplied 
by the TF slope so that it can be expressed in mags). It is clear
that the {\it intrinsic} scatter makes an important contribution to the
total error budget. Such contribution varies between $\sim 0.4$ mag
for low width objects and $\sim 0.2$ mag for high width ones. In
Figure 1, the {\it total measurement and processing error}, $\epsilon_m$,
which is the result of the combination of magnitude ($\epsilon_y$)
and width ($|b|\epsilon_x$, where $b$ is the TF slope) errors, is
indicated by a thin, solid line. $|b|\epsilon_x$ and $\epsilon_y$
are {\it not} added in quadrature, because errors in the two 
coordinates are coupled via inclination corrections.

\noindent (iii) Measurement and processing errors on the magnitude
can be important drivers of the scatter, especially for luminous,
highly inclined galaxies.

Franx and de Zeeuw (1992) have found that the TF scatter poses strong 
constraints on the elongation of the gravitational potential in the
disk plane of spirals. Their conclusion, that the average ellipticity of 
the potential in the plane of the disk must be smaller than about 0.06,
is reinforced by the scatter amplitude revealed in Figure 1. On the basis 
of 2.2 $\mu$m photometry of 18 face--on spirals, Rix and Zaritsky (1995) 
find that departures from disk axisymmetry may contribute $\sim 0.15$ mag 
to the TF scatter. Their conclusions apply principally to the inner
regions of the disk, which are sampled by the K--band observations, and
to TF scatter based on optical observations of the rotation
field. When the TF relation is based on I band photometry and 21 cm 
spectroscopy, the Rix \& Zaritsky effect becomes of more ambiguous
interpretation, as both the light and the HI emission
arise in outer regions of disks. Eisenstein
and Loeb (1996) estimate that the scatter resulting from varying formation
histories of galaxies should exceed 0.3 mag for a broad class of cosmological 
scenarios. The relatively low values found for the intrinsic scatter in 
Figure 1 suggest either an unexpectedly late epoch of galaxy formation or 
that a secular, regularizing feedback mechanism may be responsible 
for the tightness of the TF relation, as suggested by Silk (1996).

Sandage and collaborators (1994a,b; 1995) advocate for a large value of the
TF scatter, near or larger than 0.7 mag, as an explanation for the high
values of the Hubble constant resulting from the use of the TF relation. 
If the scatter were as large as proposed by that group, large biases
would result; their correction would change the zero--point of the TF 
template relation in the sense that the value of $H_\circ$ would be
reduced. While the values of the scatter shown in Figure 1 
are not as low as advocated by other groups, it appears unlikely that the 
scatter may be as large as suggested by Sandage \etal .

It has been advocated that the use of an {\it inverse} fit for the TF relation 
--- one where the ``independent'' variable is the magnitude rather than the
velocity width --- does away with the need to correct for incompleteness
bias (e.g. Schechter 1980). The nature of the TF scatter, 
especially the fact that velocity width errors can be overshadowed by
other sources, weakens the case for a bias--less inverse TF relation.

\section {TF Template Relation, Bias and Other Corrections}

The construction of a template relation is the most delicate aspect of 
any TF program. Providing a large number of objects located at a common
distance, and therefore exempt from the vagaries introduced by an {\it
a priori} unknown peculiar velocity field in a field galaxy sample, 
clusters of galaxies are favorite targets for the determination of the 
properties of the TF relation. They are, however, not exempt from the
necessity to evaluate and apply important corrections that take into
consideration the interplay between scatter and sample completeness,
the non--negligible line--of--sight extent of the cluster, corrections
for morphological type mix and others. It is moreover necessary to
verify whether a cluster's environment alters the photometric and
kinematical properties of galaxies, so that the derived TF relation
for the cluster may not be applicable to the field. In this section,
those issues are discussed principally in light of the results obtained
by G96, using a sample of galaxies in 24 clusters at $cz$ between 1,000 
and 10,000 \kms.

\subsection {Single Cluster, Basket of Clusters}

One commonly--adopted approach to the determination of a TF template 
relation is to select a single cluster of galaxies as a reference, 
thereby equating the universal template with  the TF relation defined 
by its members. There are several problems with this approach.
In order for the cluster to constrain well the TF slope, a wide dynamic
range in each of the TF parameters is desirable, which advocates for the 
use of a nearby cluster, in which less luminous galaxies can easily be
targeted. Such a cluster would, however, yield a highly uncertain
zero point, since even a modest \vpec would result in a large magnitude
offset. Conversely, a distant cluster, for which the latter problem 
would be minimized, would provide lax constraint for the TF slope. 
A single cluster sample is moreover unlikely to contain more than 
two or three dozens of objects,
which would produce a template of very poor statistical definition.
TF relations of very low scatter are sometimes found in such cases, 
which reflect far more the capriciousness of small--number statistics
than exceptional data quality. In those cases, low scatter
is seldom accompanied by accurate slope and zero--point. 

The alternative approach is that of using a set of clusters rather than 
a single one. More distant clusters in the set can effectively
provide an estimate of the velocity zero point, while the nearby ones
can help constrain the TF slope. The combination of the various data
sets from several clusters does however require the simultaneous
estimate of their relative \vpec, as well as of the bias and other 
corrections that apply differentially to each. The procedures followed
in obtaining such a combination are discussed as follows.

\subsection {The Incompleteness Bias}

Much has been written on the ways of estimating the incompleteness bias, 
which results from the interplay between the degree of completeness of a
sample and the amplitude of the TF scatter relation (e.g. Bottinelli \etal 
1986; Sandage 1994a;
Sandage \etal 1995; Teerikorpi 1993 and refs. therein). The nature of
the bias is illustrated in the simulation shown in Figure 2, where
\begin{figure} 
\epsscale{0.6}
\plotone{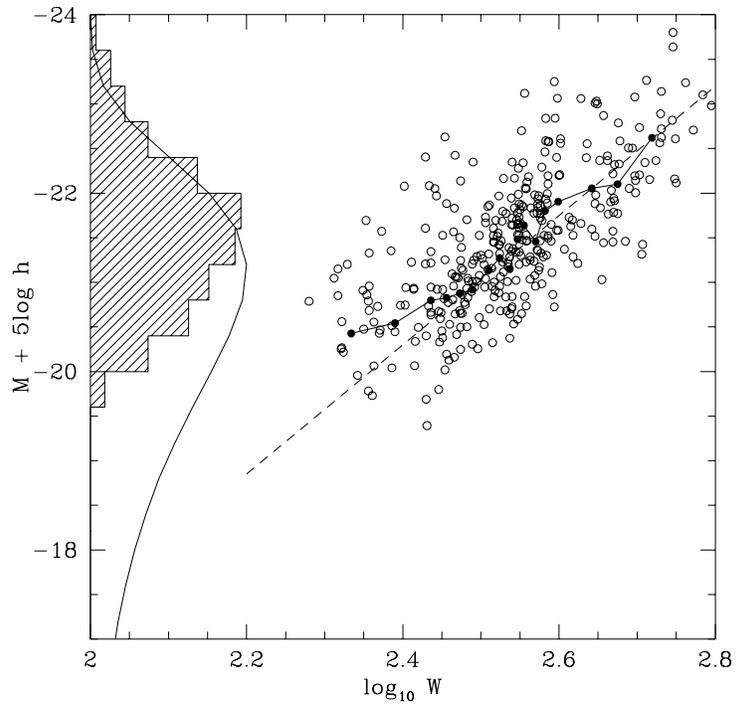}
\caption {Simulated dramatization of the incompleteness bias. See 
text for explanation.}
\end{figure}
a `cluster sample' is extracted from a population with a luminosity 
function (LF) as shown by the smooth curve plotted along the vertical 
axis. The extracted sample is however incomplete, as the histogram of 
magnitudes superimposed on the LF shows. Incompleteness exhibits a 
``soft'' edge, indicated by the progressive departure of the histogram 
from the LF. For a flux--limited sample, the histogram would of course 
track the LF for magnitudes brighter than the limit, then level suddenly 
to zero. The TF law assigned to the simulated data (UTF) is represented 
by a dashed line, with a scatter twice as large as that illustrated in 
Figure 1, averaging about 0.7 mag. The magnified scatter serves to dramatize 
the bias. A heavy solid line connects filled symbols which identify mean values 
of the magnitude within bins of velocity width. Incompleteness affects the 
TF relation derived from the simulated sample in several important ways:
(i) the derived slope is less steep than that of the UTF; (ii) the 
zero--point is brighter than that of the UTF; (iii) the scatter is
underestimated with respect to that of the UTF. 

Correction recipes for the incompleteness bias have been given in a
variety of studies, most recently those by Teerikorpi (1993 and refs.
therein), Willick (1994), Sandage \etal (1995) and G96.
Unlike most other works, which propose analytical or graphic solutions
to the problem, G96 give a solution obtained via
Monte Carlo simulations, which applies to the case
when the TF relation is obtained by fitting magnitude on the logarithm
of the velocity width, taking into simultaneous account errors on both 
coordinates. This case, referred to as the {\it bivariate} fit, differs 
fron the one usually referred to as the {\it direct} fit, in which
only errors in magnitude are taken into account. The computation 
of bias corrections does of course depend on the chosen type of fit.

The incompleteness bias generally increases with increasing
distance to the cluster. It is principally driven by the amplitude of 
the TF scatter. An important source of uncertainty in its application
is related to the shape of the LF of the galaxy population from which
the cluster sample is extracted. G96 estimate that the
effect of this uncertainty on that of the TF zero--point is 
$\sim 0.03$ mag.

\subsection {TF Dependence on Morphology and Cluster Environment}

Early work by Roberts (1978), de Vaucouleurs \etal (1982) and others
showed that in the blue part of the visible spectrum there are significant
differences in TF behavior among galaxies of different morphological types.
Aaronson and Mould (1983) found such differences to be imperceptible when
H band photometry is used. At I band, G96 find a
weak but significant difference in the zero point between early--
and late--type spirals, Sa and Sab galaxies being less luminous, at a
given width, than Sbc and Sc galaxies. Because cluster samples often
include a sizable fraction of early--type spirals, an adjustment for
the mixing needs to be made.

\begin{figure} 
\centerline{\psfig{figure=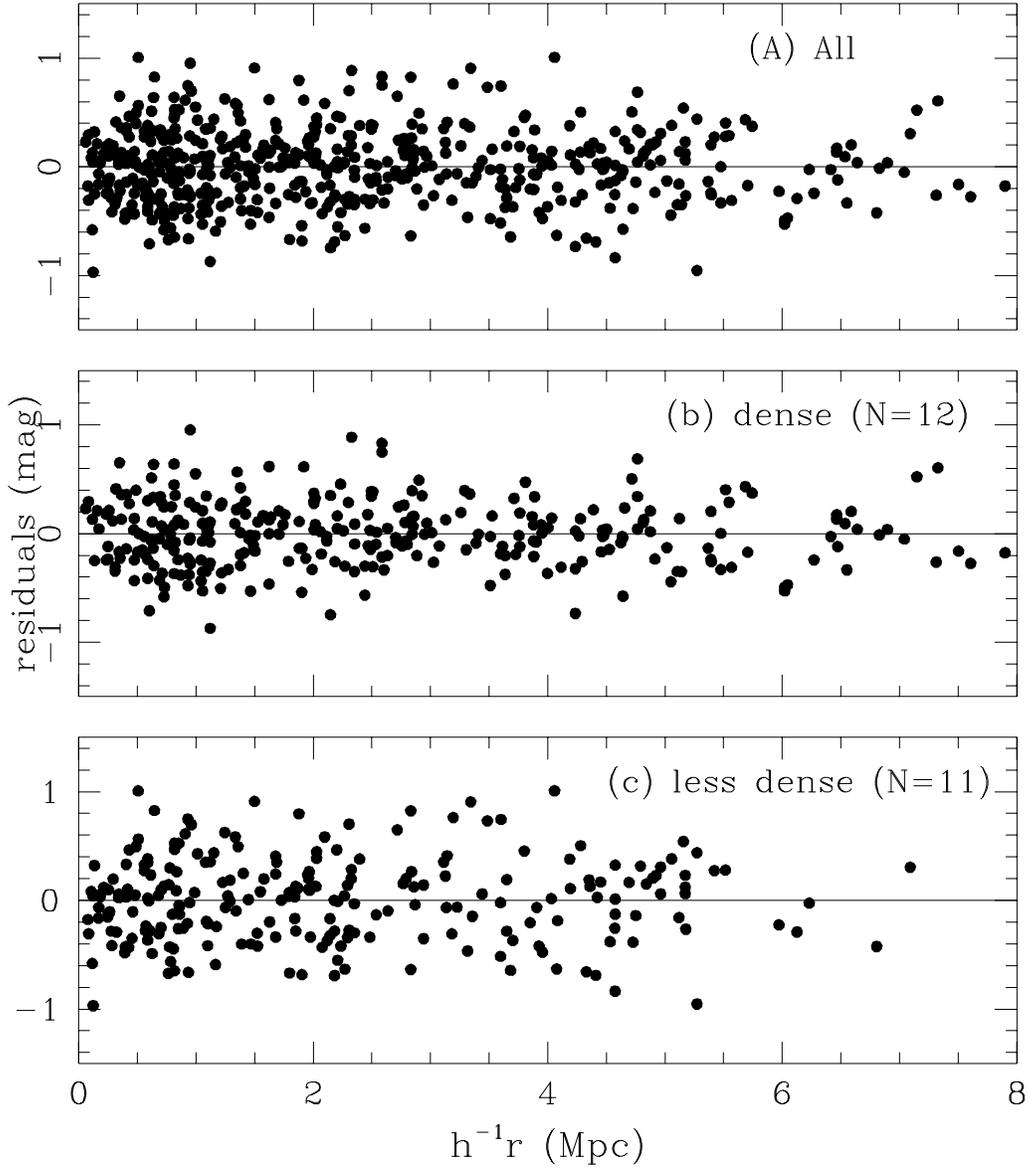,height=7.0in}}
\caption {TF residuals versus projected radial distance of galaxies
from cluster centers. In panel (a), all 555 galaxies in 24 clusters are
shown, while in panels (b) and (c) galaxies are separated according to
whether they lie in the richer or lower half of the cluster set.}
\end{figure}

In Figure 3, the residuals from the I band TF template relation are 
shown as a function of projected radial distance from the center of 
clusters. In panels (b) and (c) galaxies are separated between the high 
and low richness halves of the 24 cluster set of G96, while in panel (a) 
the total cluster sample is displayed at once. There is no significant 
evidence of differential TF behavior among spirals, on the basis of their
projected cluster location.

\subsection {The I Band TF Template Relation}

Using 14 clusters with $cz > 4,000$ \kms to define the zero--point
and the whole set of 24 clusters to constrain the slope, G96 obtain
a template relation based on the data shown in Figure 4, after the 
necessary corrections for bias, morphology, cluster extent and peculiar 
velocity have been applied. The plot includes 555 galaxies. The best
bivariate fit is 
$$M+5\log h = -21.00\pm0.02 - (7.67\pm-0.11)(\log W -2.5) \eqno(10).$$
The statistical errors on the coefficients are small, but as
we shall see they are not the principal contributors to the uncertainty
of the peculiar velocities inferred using the template.

\begin{figure} 
\centerline{\psfig{figure=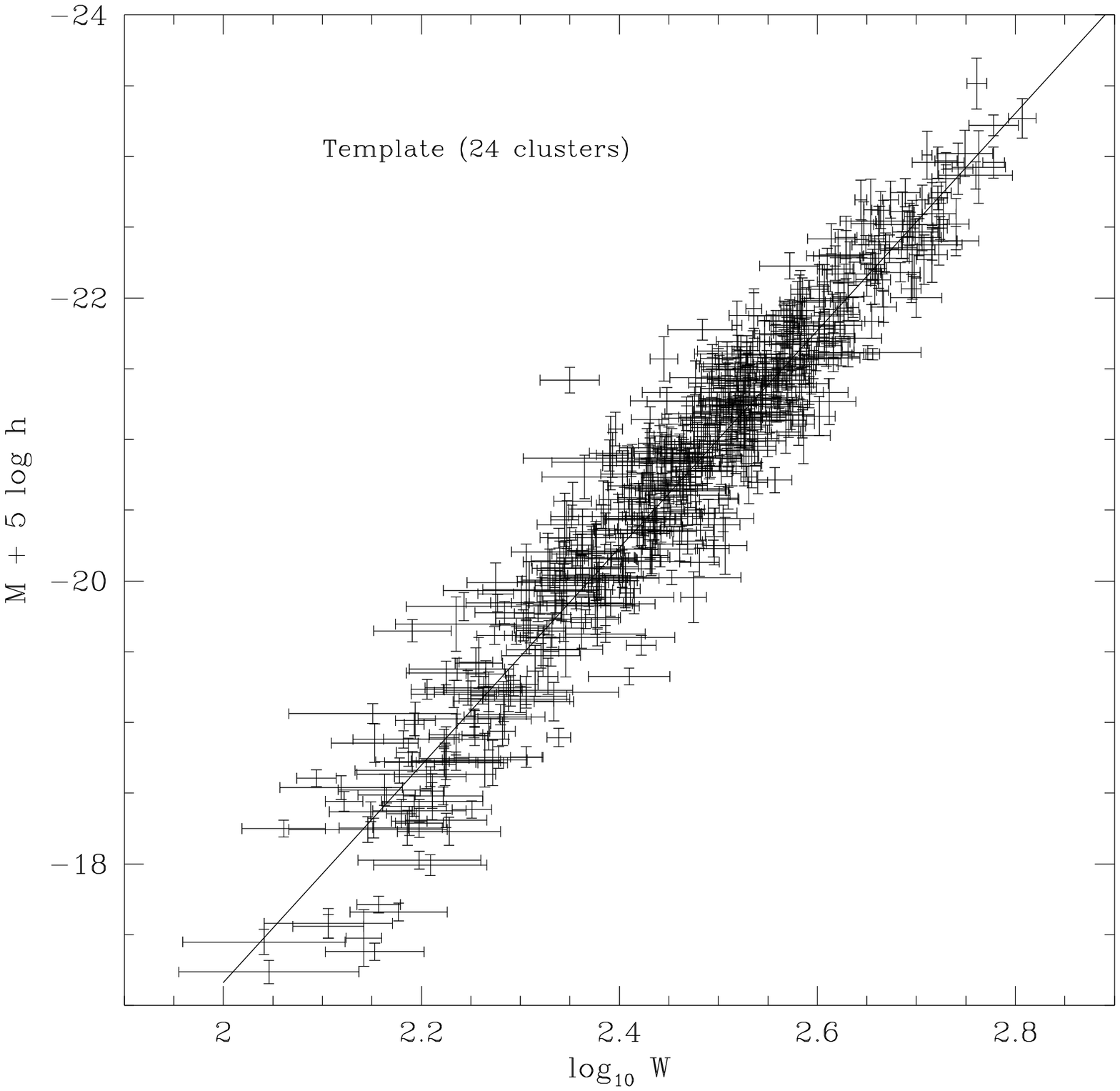,height=6.0in}}
\caption {Template relation based on 555 galaxies in 24 clusters.}
\end{figure}

\section {Absolute Calibration of the TF Relation}

The final step in the calibration of the TF relation towards deriving
an estimate of \hnot relies on the availability of primary distances 
for galaxies with TF parameters of adequate quality. Such galaxies
would be preferably fairly inclined, isolated late spirals, exhibiting
no evidence of perturbation in either their light distribution or 
velocity field. Table 1 lists galaxies with Cepheid distances, that can 
be potentially used as TF calibrators.
Names in various catalogues are listed in cols. (1) to (3) and the morphological
type in the RC3 system ('2' stands for Sab, '3' for Sb, '5' for Sc) is given
in col. (4). The adopted inclination of the disk is in col. (5), averaged
from various estimates in the literature; the raw, total I band apparent
magnitude $I_{tot}$ is listed in col. (6) and the assumed galactic extinction 
and internal extinction corrections, $A_{MW}$ and $A_{int}$, are respectively
in cols. (7) and (8). The morphological type correction $\beta_{type}$, 
necessary to make the TF behavior of galaxies earlier than Sbc consistent 
with that of Sbc and Sc objects, is given in col. (9), following the recipe
of G96. The adopted distance modulus $DM$ and its estimated error are 
in col. (10), as given in several recent publications
and as presented in this conference. The absolute magnitude, obtained via
$$M_I = I_{tot} - A_{MW} - A_{int} - \beta_{typ} - DM \eqno (11)$$
is in col. (11). Its uncertainty, in brackets on the last two significant figures
of $M_I$, has been computed assuming a magnitude measurement uncertainty of 0.1 
mag and propagating errors on the extinction correction. The adopted value 
of the logarithm of the velocity width $W$ and its estimated error are listed 
in col. (12). The width corresponds to approximate measurements at 50\% HI peak 
flux intensity, after correction for resolution, turbulent motions and 
inclination effects.

\begin{deluxetable}{lrrrrrrrrrrr}
\tablewidth{0pt}
\tablenum{1}
\tablecaption{TF Primary Distance Calibrators: Adopted Parameters}
\tablehead{
\colhead{UGC}             & \colhead{NGC}            & \colhead{M}              & 
\colhead{T}            & \colhead{$i$}      & \colhead{$I_{tot}$}      & 
\colhead{$A_{gal}$}       & \colhead{$A_{int}$}      &
\colhead{$\beta_{typ}$}      & \colhead{DM}             & 
\colhead{M$_I$}           &  
\colhead{$\log W$}          
\nl
\tablevspace{5pt}
\colhead{(1)} & \colhead{(2)} & \colhead{(3)} & \colhead{(4)} & \colhead{(5)} &
\colhead{(6)} & \colhead{(7)} & \colhead{(8)} & \colhead{(9)} & \colhead{(10)}&
\colhead{(11)}& \colhead{(12)}
}
\startdata
   454  &  224  &  31 &  3 & 78 & 2.18 & 0.14 & 0.65 & 0.10 & 24.44(12) &  -23.15(19) &   2.663(39) \nl
 20647  &  300  &     &  8 & 43 & 7.36 & 0.02 & 0.09 & 0.00 & 26.66(15) &  -19.43(18) &   2.269(33) \nl
  1117  &  598  &  33 &  5 & 57 & 4.98 & 0.07 & 0.20 & 0.00 & 24.64(10) &  -19.92(17) &   2.286(32) \nl
  1913  &  925  &     &  5 & 57 & 9.30 & 0.11 & 0.26 & 0.00 & 29.86(16) &  -20.93(18) &   2.330(46) \nl
 22699  & 1365  &     &  3 & 48 & 8.35 & 0.00 & 0.11 & 0.10 & 31.30(17) &  -23.16(19) &   2.684(15) \nl
  3851  & 2366  &     &  8 & 69 &11.05 & 0.07 & 0.22 & 0.00 & 27.68(20) &  -16.92(22) &   1.887(77) \nl
  3918  & 2403  &     &  5 & 59 & 7.39 & 0.06 & 0.25 & 0.00 & 27.51(15) &  -20.43(18) &   2.389(42) \nl
  5318  & 3031  &  81 &  2 & 57 & 5.70 & 0.08 & 0.28 & 0.32 & 27.80(20) &  -22.78(18) &   2.602(55) \nl
 26815  & 3109  &     & 10 & 80 & 9.23 & 0.06 & 0.35 & 0.00 & 25.45(20) &  -16.63(22) &   2.079(43) \nl
  5882  & 3368  &  96 &  2 & 42 & 8.10 & 0.03 & 0.15 & 0.32 & 30.32(16) &  -22.72(19) &   2.633(34) \nl
  7450  & 4321  & 100 &  5 & 28 & 8.22 & 0.02 & 0.06 & 0.00 & 31.04(15) &  -22.91(16) &   2.681(63) \nl
  7668  & 4496  &     &  5 & 34 &10.88 & 0.00 & 0.08 & 0.00 & 31.13(13) &  -20.33(16) &   2.412(80) \nl
  7732  & 4536  &     &  4 & 70 & 9.53 & 0.01 & 0.45 & 0.00 & 31.11(13) &  -22.04(18) &   2.489(16) \nl
  7884  & 4639  &     &  3 & 55 &10.49 & 0.02 & 0.24 & 0.10 & 32.00(23) &  -21.77(25) &   2.534(25) \nl
  8981  & 5457  & 101 &  5 & 21 & 6.97 & 0.00 & 0.04 & 0.00 & 29.34(17) &  -22.37(15) &   2.560(77) \nl
\enddata
\end{deluxetable}

The raw I band magnitudes are from Pierce and Tully (1992) and Pierce 
(1996), except for NGC 1365, for which we use the average of the Mathewson
\etal (1992) and of the Bureau \etal (1996) values. The spectrospcopic 
data are from a wide variety of sources, too long to itemize here. 
The velocity widths of large, nearby objects pose accuracy problems of  
singular nature. Some the data are generally old and unavailable in 
digital form. In several of the objects, clear distortions are 
discernible: warps, tidal perturbations and other asymmetries reduce
the reliability for TF calibration (e.g. M31, M33 and M81). Other
systems are dwarf irregulars (e.g. NGC 2366 and NGC 3109), and thus
ill-suited for use with the G96 template which is principally constructed 
using luminous spirals. Two galaxies (M100 and M101) have very low 
inclinations and require large and uncertain corrections to the observed
widths. One system (NGC 4496) is not only nearly face--on, but a second
galaxy is seen superimposed on its disk, making the extraction of
photometric parameters highly uncertain. In practice, our estimates of TF 
parameters are based on a somewhat subjective synthesis of a large amount 
of heterogeneous material, sometimes involving the measurement of spectra on 
paper copies of published data figures. Our assignment of error bars to the 
data are thus reflective of this unorthodox method of parameter derivation, 
rather than of the original accuracy of the data.

\section {The Value of H$_\circ$}

In Figure 5, the data of all galaxies in table 1 are plotted over a grid 
of renditions of the template relation, derived as discussed in section 4.4 
and shifted according to values of the Hubble parameter ranging
between $h=0.5$ and $h=1.0$. Keeping the slope of the TF template fixed, 
we compute the value of $h$ that yields $\chi^2$ minimization of residuals 
for the set of calibrators. The statistical uncertainty of the derivation, 
based on the assigned errors to the velocity 
widths and absolute magnitudes of the calibrators, is small. 
If we use all the galaxies listed in Table 1, the resulting best fit value 
is $h = 0.74\pm 0.02$.
However, several of the objects listed in Table 1 are very ill--suited for
calibration, as discussed above. After exclusion of N4496 (interacting pair
with companion superimposed), N2366 and N3109 (Irregular, low luminosity 
types), which are identified by unfilled symbols in Figure 5, the best fit 
yields
$$h = 0.70\pm 0.02 \eqno(12)$$
The formal errors in the statistical analysis given above, of $\sim 0.06$ 
mag, arise purely from the errors on the TF parameters of the calibrators,
which are likely to be underestimated. Very large galaxies, for example, 
pose difficulties in the estimation of magnitudes because of problems 
associated with the determination of the sky brightness level, and the
assumed 0.1 mag measurement error may in some cases be too small. Corrections 
to widths, necessary to bring them to an internally compatible system, as 
well as commensurable with the data used to obtain the template relation, 
are also quite uncertain, when the data originate in as heterogenous a set 
of data sources as in this case. A realistic appraisal of the error 
arising from measurements and corrections applied to the TF parameters 
may be significantly larger than the statistical estimates given above. 
More conservatively than in Table 1, we shall assume that uncertainties 
in the calibrators' TF parameters 
contribute $\Delta m_{cal} \sim 0.10$ mag to the error budget.

The error $\Delta m_{cal}$ does not include the possibility of systematic 
bias in the zero point of the Cepheid period--luminosity relation and
in the determination of the coefficients of the TF template. We deal
with the latter uncertainty next.

\begin{figure} 
\centerline{\psfig{figure=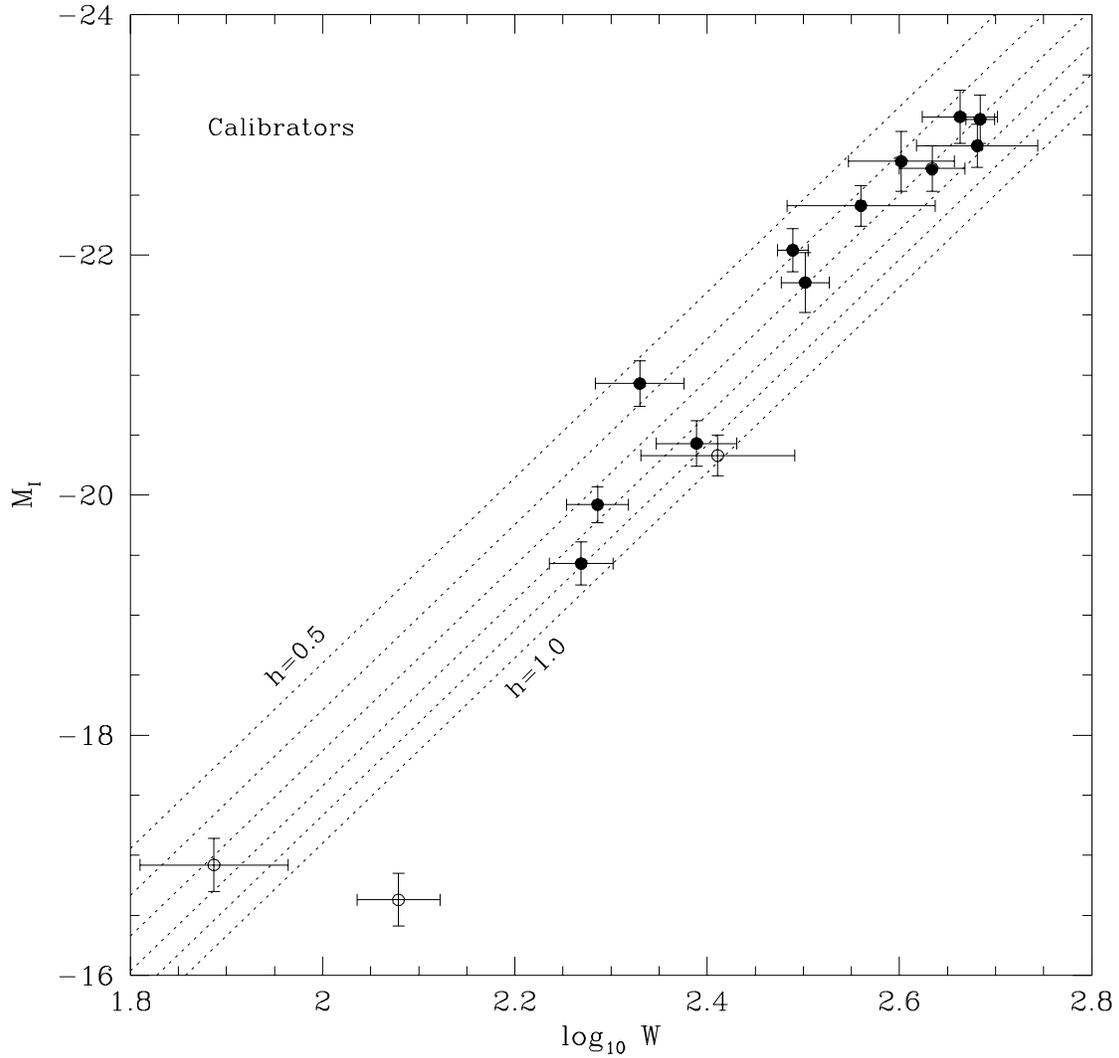,height=6.0in}}
\caption {TF calibrators with Cepheid distances, superimposed on
a grid of TF template relations plotted for different values of $h$.
The unfilled symbols represent galaxies unfit for TF work, as 
explained in the text.}
\end{figure}

The statistical accuracy of the TF zero point deriving from the scatter
of the 555 data points about the template shown in Figure 4, is $\sim 0.02$
mag,as given in section 4.4. In addition, two other sources of uncertainty
need to be considered: (a) that resulting from the 
choice of an LF for the computation of the incompleteness bias corrections, 
which in section 4.2 was estimated to contribute 0.03 mag, and (b) that 
resulting from the assumption that the set of clusters at $cz>4000$ \kms 
defines a system with $<V_{pec}=0>$. Part (b) can be estimated as follows.
For $N$ randomly selected clusters, 
located at a mean systemic velocity $<cz>$ and characterized by a r.m.s. 
1--d peculiar velocity $<V_{pec}^2>^{1/2}$, the magnitude error on the 
TF zero point obtained by assuming that the set of clusters yields 
$<V_{pec}=0>$ is
$$\Delta m \sim 2.17 <V_{pec}^2>^{1/2} <cz>^{-1} N^{-1/2} \eqno (13)$$
The value of $<V_{pec}^2>^{1/2}$ is quite uncertain. Taking a value
intermediate between that suggested by the data of G96 and that required
by flat CDM cosmological models (see discussions by Bahcall and Oh 1996
and Moscardini \etal 1996), and allowing for a measure of correlation
in cluster locations, we obtain $\Delta m \sim 0.06$ mag. The combined
uncertainty on the TF zero point is thus $\Delta m_{zp} = (0.06^2 + 
0.03^2)^{1/2} \simeq 0.07$ mag.

Combining now in quadrature $\Delta m_{cal}\sim 0.10$ mag, 
$\Delta m_{zp} \sim 0.07$ and an arbitrarily assumed uncertainty on the
Cepheid P--L relation of 0.1 mag, we obtained a total uncertainty of 
0.16 mag which, updating expression (12), yields
$$H_\circ = 70\pm 5. \eqno(14)$$

\section {The Peculiar Velocities of the Virgo and Fornax Clusters}

The peculiar velocities of Virgo and Fornax are of particular interest
since several of the most recent Cepheid distance determinations 
correspond to galaxies thought to be members of those clusters.
While the route to estimating a value of \hnot via the TF template
relation followed in the preceding section is more accurate, it is
useful to have at hand estimates of the \vpec of the two clusters, which
in combination with their systemic velocities and distances
can yield local estimates of \hnot. Note that the \vpec given here are
referred to the same TF template relation used in the preceding section; 
thus the inferred values of \hnot given below are not fully independent on
that given by (14).

Fornax is part of the cluster set of G96. Its \vpec, based
on a TF sample of 26 galaxies thought to be cluster members and measured
in the CMB reference frame, is $-26\pm71$ \kms. By expanding the sample
to include 13 additional objects thought to be in the cluster periphery,
one obtains $V_{pec} = -103\pm66$ \kms. Both values assume a systemic
recesssion velocity for the cluster of $1321\pm45$ \kms, measured
again in the CMB reference frame. If Fornax lies at the distance
of NGC 1365, of $18.2\pm1.5$ Mpc (Silbermann \etal 1996), and if its
Hubble velocity $H_\circ d = 1385$ \kms, then \hnot $= 76\pm8$ \kmsm.

Virgo is not one of the clusters studied by G96. We have, however,
used the I band data of Pierce and Tully (1992) and of Pierce (1988),
and selected a sample of galaxies reputed to be members of the `A'
clump, according to the criteria of Binggeli \etal (1985; 1993).
Clump A is centered at the position of M87 and has a systemic velocity  
of $1378\pm35$ \kms (equivalent to a heliocentric velocity of 
of 1050 \kms as found by Binggeli \etal 1993). Twentythree galaxies are
used to obtain $V_{pec} = 204\pm65$ \kms, following procedures analogous
to those described in G96. If Virgo lies at the mean distance of M100, 
N4496 and N4639 of $17.4\pm 0.6$ Mpc, then \hnot $= 67\pm8$ \kmsm.

\acknowledgements
It is a pleasure to thank Dr. Michael Pierce for generously providing 
data in advance of publication, help in obtaining the latest Cepheid
distances and enjoyable conversations, as well as Dr. M. S. Roberts for 
providing insights on the velocity field of M31.
This research was partially supported by NSF grant AST94--20505.

\vfill
\end{document}